\documentclass[pdflatex,sn-mathphys-num]{sn-jnl}
\usepackage{graphicx}%
\usepackage{multirow}%
\usepackage{amsmath,amssymb,amsfonts}%
\usepackage{amsthm}%
\usepackage{mathrsfs}%
\usepackage[title]{appendix}%
\usepackage{xcolor}%
\usepackage{textcomp}%
\usepackage{manyfoot}%
\usepackage{booktabs}%
\usepackage{algorithm}%
\usepackage{algorithmicx}%
\usepackage{algpseudocode}%
\usepackage{listings}%

\usepackage{subfigure}
\usepackage{anyfontsize}

\theoremstyle{thmstyleone}%

\theoremstyle{thmstyletwo}%

\theoremstyle{thmstylethree}%

\begin{document}

\title[Article Title]{A 2048-spin bulk acoustic wave Ising machine for number partitioning and Sudoku}



\author*[1]{Venkatesh Vadde} \email{venkatesh.vadde@physics.gu.se}
\author[1]{Roman Ovcharov}
\author[1]{Victor H. González}
\author[1]{Roman Khymyn}
\author*[1,2]{Artem Litvinenko} \email{litvinenko@oakland.edu}
\author*[1,3,4]{Johan \AA kerman} \email{johan.akerman@physics.gu.se}

\affil[1]{Department of Physics, University of Gothenburg, Gothenburg, Sweden}
\affil[2]{Department of Physics, Oakland University, Rochester, Michigan, USA}
\affil[3]{Center for Science and Innovation in Spintronics, Tohoku University, Sendai, Japan}
\affil[4]{Research Institute of Electrical Communication, Tohoku University, Sendai, Japan}

\abstract{
Optical coherent Ising machines based on time-multiplexing have demonstrated significant progress in terms of connectivity and spin scalability. However, they are constrained by large physical footprints, high power consumption, poor thermal stability, and high cost. 
Here, we present a time-multiplexed Ising machine leveraging propagating wave packets in solid-state delay lines at microwave frequencies, enabling thermally stable, robust, low-power, tabletop, and affordable design. We use two serially connected 20.5 MHz, 707 $\mu s$ bulk acoustic wave delay lines supporting 2,048 spins. Our design provides all-to-all connectivity with 15-bit coupling resolution and finds approximate MAX-CUT solutions in 341 ms, potentially scalable to sub-ms by using higher frequency delay lines.
Additionally, we demonstrate solutions to number partitioning and Sudoku problems. Compared with state-of-the-art Coherent Ising machines, our machine exhibits four orders of magnitude higher thermal stability. Against the simulated bifurcation algorithm, our design achieves comparable results on the MAX-CUT problem, while outperforming it on the more complex number-partitioning and Sudoku problems.
}
\maketitle

\section{Introduction}\label{sec1}

The slowing of Moore’s law in terms of advancement in modern computers appears to have reached a plateau, promoting the exploration of physics-based unconventional computational architectures beyond traditional silicon-based technology. This shift is particularly important for addressing non-deterministic polynomial-time hard (NP-hard) and NP-complete problems, which represent classes of highly complex challenges, especially in combinatorial optimization, where solution times increase exponentially with problem size, posing significant difficulties for classical computing approaches. These optimization problems~\cite{barahona1982computational,du1998handbook} play a crucial role in a wide range of fields, including finance~\cite{ibarra1975fast}, circuit design~\cite{barahona1988application}, drug discovery~\cite{earl2005parallel}, operations~\cite{vcerny1985thermodynamical}, and scheduling~\cite{burke2004state}. 
Combinatorial optimization problems are known to be mappable onto ground-state search problems of the Ising model using polynomial resources~\cite{barahona1982computational}. 

The Ising Hamiltonian is given by, 
\begin{equation}
    H = - \sum_{i<j}^{}  J_{ij} s_i s_j - \sum_{i} h_i s_i 
    \label{eqn:hamiltonian}
\end{equation}
where $s_i = \pm 1$ corresponds to the $i^{th}$ Ising spin, $J_{ij}$ is the coupling term between spins $s_i$ and $s_j$, and $h_i$ is a local bias field. The Ising machines are engineered to find configurations minimizing this Hamiltonian.

To address these computational demands of NP-hard problems, novel approaches have emerged that utilize the physical behavior of systems to perform efficient computation, such as analog Ising machines.
A wide range of Ising machines have been developed using diverse physical platforms, including superconducting quantum bits~\cite{johnson2011quantum}, single-electron devices~\cite{nishiguchi2007single}, nanomechanical systems~\cite{mahboob2016electromechanical}, stochastic nanomagnets~\cite{sutton2017intrinsic}, CMOS circuits~\cite{whitehead2023cmos}, spin waves~\cite{litvinenko2023spinwave, gonzalez2024global, gonzalez2024spintronic}, superparamagnetic tunnel junctions~\cite{si2024energy}, delay-line oscillators~\cite{ovcharov2024numerical}, and photonic Ising machines~\cite{yamamoto2017coherent,honjo2021100}. Several commercial products are also available on the market for solving combinatorial problems, including the D-Wave system, which uses superconducting technology, NTT Research’s optical Coherent Ising machine, Fujitsu’s Digital Annealer, and Hitachi’s CMOS Annealing Machine. However, each of these devices has limitations, such as the need for cryogenic temperatures, high power consumption, and confinement to laboratory environments.

Analog Ising machines using time-multiplexing have seen significant advancements, beginning with the demonstration of the optical Coherent Ising Machine (CIM)~\cite{takesue2025finding,honjo2021100,inagaki2016coherent}, this architecture overcame connectivity limitations by enabling all-to-all coupling between spins. CIMs have been successfully scaled from as few as 4 spins to over 100,000 spins. However, despite these achievements, they remain confined to laboratory settings due to their large physical footprint, high power consumption, poor temperature stability, and high cost. Some of these challenges have been partially addressed by the Spinwave Ising Machine (SWIM)~\cite{litvinenko2023spinwave} and surface acoustic wave-based Ising machines (SAWIM)~\cite{litvinenko202550}, where spins are represented by spinwave or acoustic wave pulses propagating at much slower speeds, on the order of kilometers per second. However, these approaches have so far only demonstrated systems with a limited number of spins---8 in SWIM and 50 in SAWIM---and are restricted to 1-bit resolution in their coupling terms.

In this work, we introduce a time-multiplexed Ising machine based on a bulk acoustic wave delay line capable of hosting 2,048 spins. Bulk-acoustic-wave-based Ising machines (BAWIMs) offer a promising approach to building thermally stable, time-multiplexed computing systems by leveraging propagating wave packets in solid-state delay lines. Compared to state-of-the-art CIMs, BAWIM is more stable and reliable, 
while eliminating the need for complex frequency-stabilization systems because their thermal stability is approximately four orders of magnitude higher.
We evaluate our Ising machine using arbitrary graphs from the BiqMac library and demonstrate that it achieves competitive energy results compared to state-of-the-art algorithmic approaches while effectively relaxing to low-energy states. 
Additionally, we report results on more complex, real-world problems, such as number partitioning and Sudoku puzzles. We show that BAWIM successfully solves the Sudoku puzzle and produces good approximate solutions for the number partitioning problem. Furthermore, we compare its performance with the simulated bifurcation algorithm and demonstrate that BAWIM outperforms it.

\begin{figure}[!ht]
    \centering
    \includegraphics[width=0.79\linewidth]{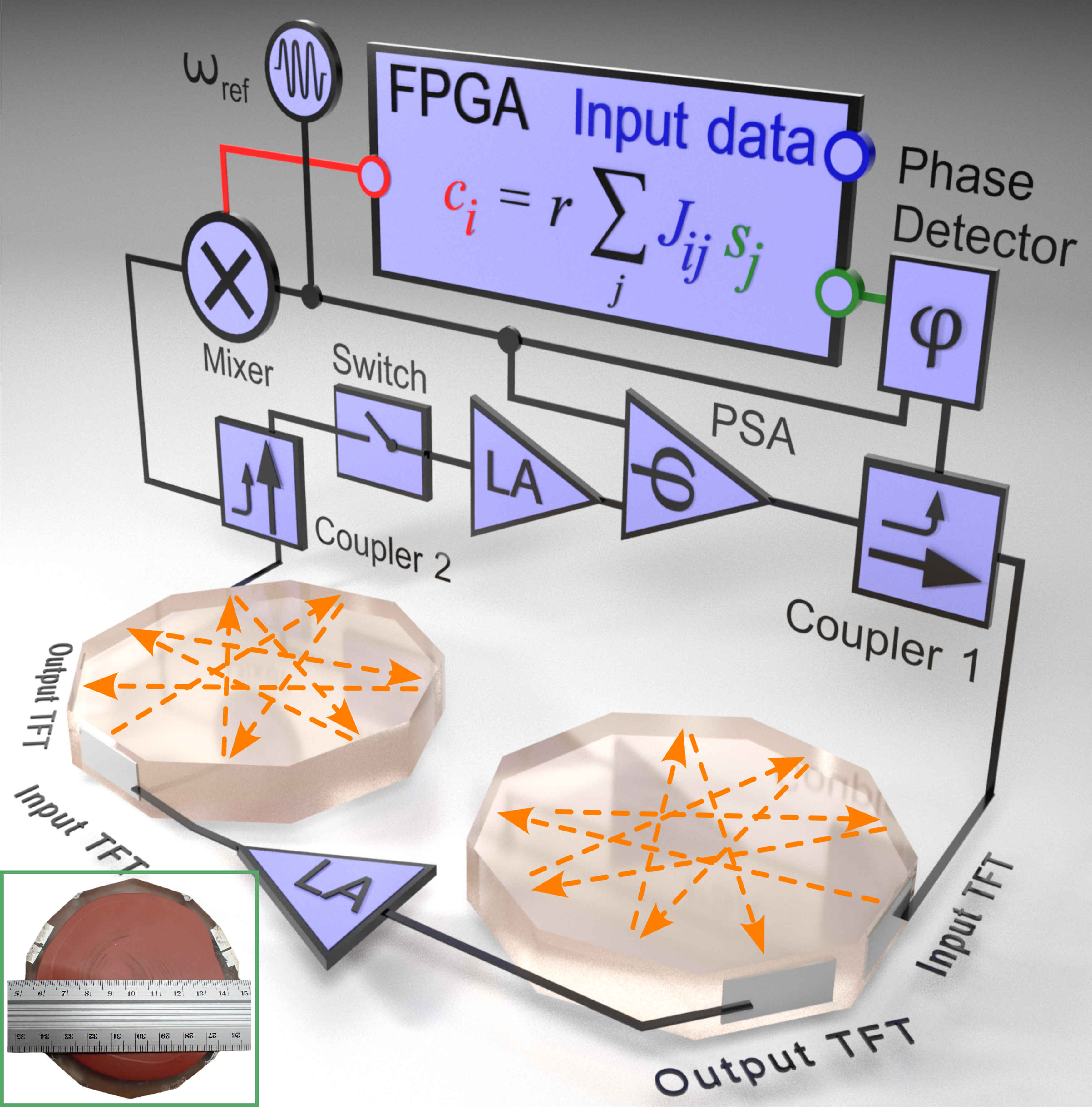}
\caption{\textbf{The Bulk acoustic wave based Ising machine (BAWIM).}
BAWIM architecture with two daisy‑chained quartz BAW delay lines for time‑multiplexed RF pulse circulation. The PSA is a parametric phase-sensitive amplifier, and the LA blocks are microwave linear amplifiers. The PSA enforces binary phases, while a measurement‑and‑feedback path reads pulse phases and injects couplings computed by the FPGA. An input and output thin-film transducer (TFT) is used to excite and receive propagating bulk acoustic waves. The inset shows the picture of the delay line.} 
\label{fig:Schema_chara} 
\end{figure}

\section{Results and discussion}\label{sec2}

\subsection{Bulk acoustic waves}
Bulk acoustic waves (BAWs) represent mechanical oscillations that propagate through the entirety of a material's volume, positioning them as highly effective carriers for signal transduction in applications such as RF filtering, sensing, and delay lines within electronic systems~\cite{hashimoto2009rf, liu2020materials}. In contrast to surface acoustic waves (SAWs), which are restricted to the substrate surface and thereby exhibit heightened vulnerability to environmental contaminants, temperature variations, and fabrication defects that can impair signal integrity~\cite{mandal2022surface, tang2024review}, BAWs confer multiple advantages, including enhanced power handling due to volumetric energy distribution, enabling high-frequency operation often exceeding 10 GHz with reduced insertion losses and elevated Q-factors for superior frequency selectivity~\cite{liu2020materials}. Furthermore, BAW devices demonstrate improved robustness against surface anomalies, better thermal stability through mechanisms like temperature-compensated designs, and greater miniaturization potential via thin-film integration~\cite{terzieva2016overview}, facilitating reliable performance in compact, high-demand scenarios such as 5G mobile communications~\cite{aigner2018baw} where maintaining signal fidelity in dense spectral environments is essential.

\begin{figure}[!hb]
    \centering
    \includegraphics[width=0.99\linewidth]{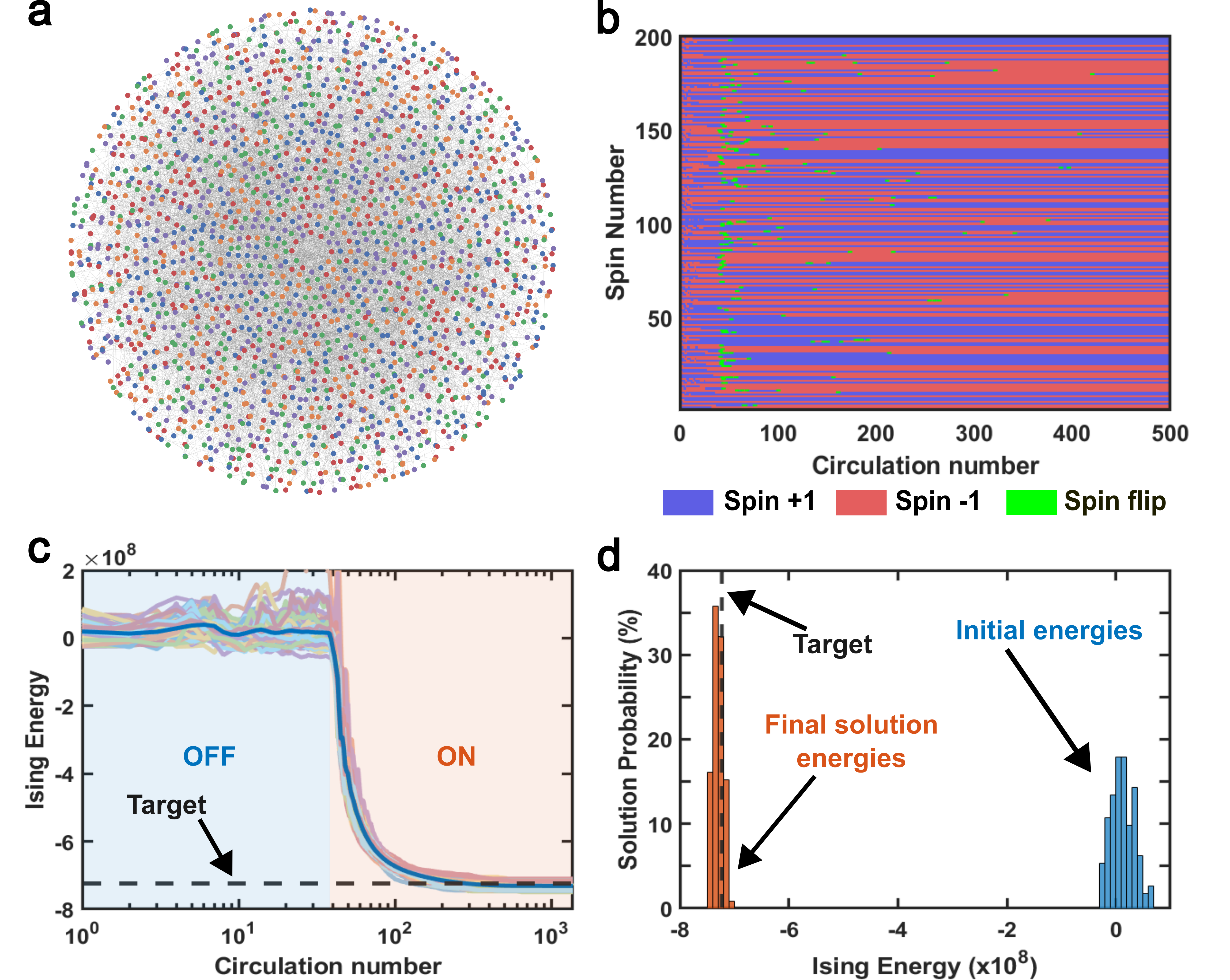}
\caption{\textbf{Experimental demonstration of BAWIM}
\textbf{(a)} Graph of the solved MAX-CUT instance with 10$\%$ edge density, comprising 209,610 edges, each with a weight between 0 and $2^{15}$; only 2,096 edges (1\%) are shown for clarity.
\textbf{(b)} Time traces of the first 200 spins during the solution run. After the 40th circulation, when the coupling is turned on, spin flips are highlighted in green.
\textbf{(c)} Time evolution of the Ising energy across all runs, with the highlighted darker blue curve showing the average solution. During the first 40 circulations, the MFB is turned off to allow the spins to settle into random states. At the 40th circulation, the MFB is switched on, initiating the solution process. The dashed line indicates the target Ising-energy, set to 90$\%$ of the best Ising energy obtained from simulated bifurcation.
\textbf{(d)} Solution probabilities for the MAX-CUT problem, showing the Ising energy before the solution process begins and the final Ising energies after completion. The dashed line shows the target Ising energy that is $90\%$ of the simulated bifurcation's best result.} 
\label{fig:Dens10_all} 
\end{figure}

\subsection{Design of a Bulk Acoustic Wave Ising Machine}
Figure \ref{fig:Schema_chara} shows the schematic of the
BAWIM, inspired by the time-multiplexing method in CIMs~\cite{honjo2021100, inagaki2016coherent}, SWIMs~\cite{litvinenko2023spinwave}, and SAWIMs ~\cite{litvinenko202550}. 
At the core of the design is a quartz-based bulk acoustic wave 707 $\mu s$ delay line, with a 3-dB bandwidth of 8 MHz centered around 20.5 MHz and with a negligible dispersion delay $\tau _{p, disp}$ of 50 ns. 
Due to the resonance mechanism underlying phase-sensitive amplification, a minimum of 5–10 cycles per pulse is required, setting a lower bound on the pulse duration to approximately 250–500 ns. This constraint dominates and effectively determines the pulse period. 
We chose a pulse period of 665 ns with a 50$\%$ duty cycle, corresponding to 6–7 cycles of the 20.5 MHz carrier signal per pulse. This is enforced in our design through a Mini-Circuits ZASWA-2-50DRA+ switch, shown in Fig \ref{fig:Schema_chara}. The spin states +1 and -1 are encoded in the relative phase of the pulses, 
with bistable phases achieved by phase-binarizing the propagating RF pulses using a phase-sensitive amplifier (PSA).
The PSA comprises a phase-sensitive attenuator combined with a linear amplifier, implemented using a 
Mini-Circuits ZYSWA-2-50DR+ device, which behaves as a 0/1 time-gate, so the signal-component at the output is largest when the gate windows line up with the signal peaks ($\phi = 0 ^\circ, 180 ^\circ $) and is strongly suppressed when they line up near the zero crossings ($\phi = \pm 90 ^\circ$). 
In addition to the components mentioned above, the BAWIM uses an Analog Devices AD835 multiplier as the mixer, an Analog Devices AD8302 as the phase detector, and Mini-Circuits Z99SC-62-S+ devices as the couplers.
As a single BAW delay line supports up to 1062 spins with a pulse width of 665 ns, we daisy-chain two delay lines, resulting in a total of  1.41 ms of delay, accommodating a total of 2124 spins. Of these, 2048 pulses are utilized as Ising spins, while the remaining 76 pulses are left unused.

The BAW delay line, together with linear amplifiers and the PSA, forms a multi-physics ring oscillator. The key step of the design is to establish a stable circulation of RF pulses within this loop. This requires satisfying the Barkhausen stability criteria: the loop gain must be larger than one, and the total phase shift around the loop must be an integer multiple of $2\pi$. The delay line introduces 36 dB of attenuation, which can be compensated by the linear amplifiers, and the phase accumulation can be ensured with a proper choice of reference signal, thus satisfying the Barkhausen criteria. The PSA further narrows the stability criteria to only those signals with either phase 0 or $\pi$ relative to the pumping signal that is twice the reference frequency.

To realize the Ising machine, we developed a measurement and feedback block (MFB) similar to
the CIMs ~\cite{honjo2021100, inagaki2016coherent}. Here, the portions of the circulating spin pulses are split with a 1:10 coupler, and their phase relative to a reference signal is measured by an AD8302 phase detector. The resulting signal is digitized by an 8-bit AD9280 ADC and processed by an AMD Zynq ZCU104 FPGA. The FPGA output is then converted back to analog using an AD9708 DAC, multiplied with the reference signal using an AD835 chip, and reinjected into the corresponding spin via a coupler.

The FPGA calculates the coupling pulses ($c_i$) based on the coupling matrix $J_{ij}$, and spins $s_j$,

\begin{equation}
     c_i = \sum_{j=1}^{N}  J_{ij} s_j  	
\end{equation}
Here the coupling terms $J_{ij}$ can have 15-bit resolution (0 to $+2^{15}$). The $c_i$ controls the phase and amplitude of the signal being injected to the circulating RF signal. The amplitude of the coupling signal can be tuned using the potentiometer of the DAC, allowing an additional control to ensure the injected signal remains within $5-30\%$ of the circulating RF signal and does not overpower it, avoiding chaotization of the oscillatory circuit.

\subsection{Performance of the Ising Machine}
We evaluate the BAWIM performance 
using MAX-CUT problems on graph instances from the BiqMac library \cite{rudy_gen, helmberg2000spectral}.
Figure \ref{fig:Dens10_all}(a) shows a 2048-node graph with an edge density of 10$\%$ having 209,610 edges, where each edge can have a weight ranging from 0 to $2^{15}$. The time evolution of the first 200 spins is presented in Fig. \ref{fig:Dens10_all}(b), where it is observed that the BAWIM immediately begins searching for lower energy states and gradually slows down as it reaches better solutions.
Figure \ref{fig:Dens10_all}(c) shows the temporal evolution of the Ising energy for multiple runs; the similar energy evolution over different runs demonstrates the reliability of our design, and it also indicates the target Ising energy based on the simulated bifurcation algorithm (details in the next section). For consecutive runs, we turn off and on the amplification in the loop with a $3\%$ duty cycle control signal with a repetition period of 2 seconds, which allows the BAW pulse to decay completely.
The quantitative results of the MAX-CUT problem are illustrated in Fig. \ref{fig:Dens10_all}(d) by a histogram of the Ising energies and MAX-CUT scores for over 100 runs, along with the target Ising energy. The performance of BAWIM depends on the coupling strength, and there exists an optimal coupling strength depending on the problem instance.

For a MAX-CUT problem, the number of cuts is given by
\begin{equation}
    Cuts = -\frac{1}{2} \sum_{i<j}^{}  J_{ij} - \frac{1}{2} H,
\end{equation}
where H is the Hamiltonian (Ising energy) in Eq. \ref{eqn:hamiltonian}.

\begin{figure}[!hb]
    \centering
    \includegraphics[width=0.99\linewidth]{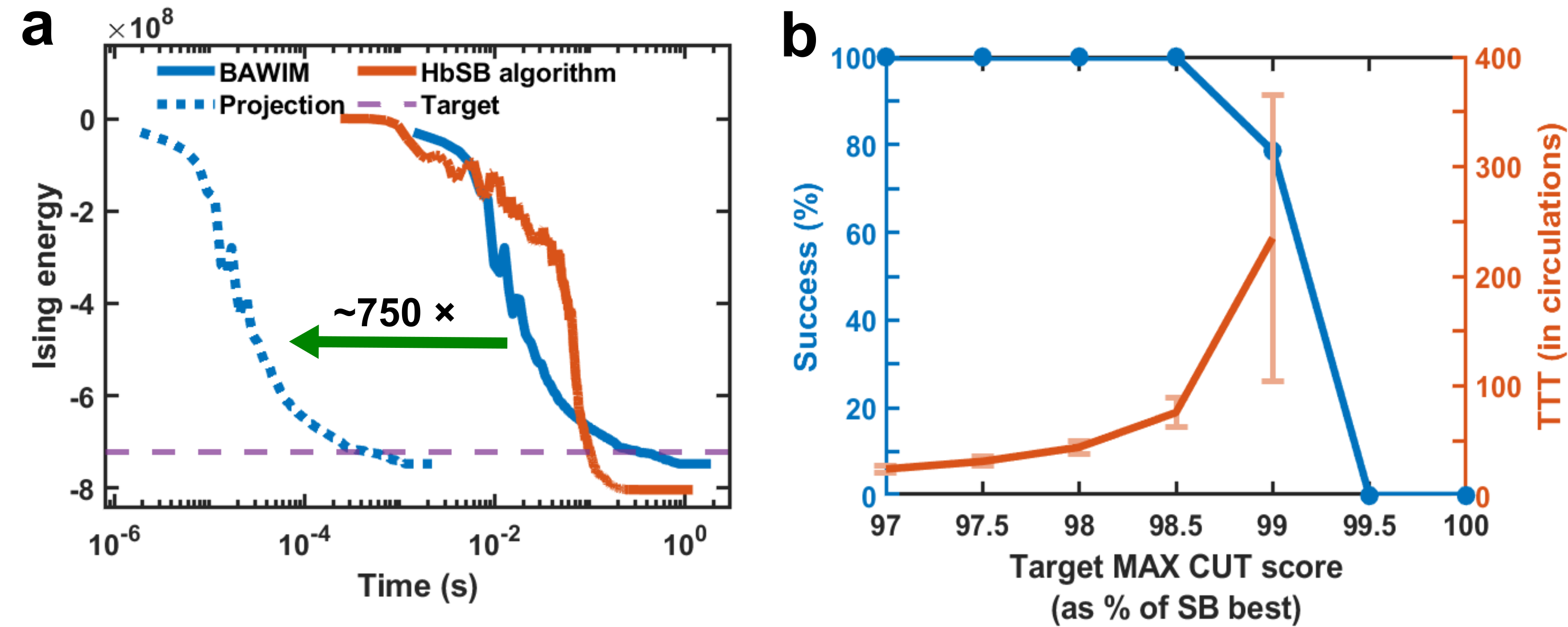}
\caption{\textbf{Comparison with heated-ballistic simulated bifurcation algorithm}
\textbf{(a)} Time evolution of the Ising energy for BAWIM, HbSB, and the projected performance of BAWIM with a high central frequency delay line, with a target Ising energy of $90 \%$ of HbSB's best result.
\textbf{(b)} Success rate and time-to-target (TTT) of BAWIM relative to the best MAX-CUT score of HbSB.
The blue curve shows the success rate computed from n = 110 independent BAWIM runs. The orange curve shows the mean TTT computed over successful runs, and the error bars indicate $\pm1$ standard deviation.}
\label{fig:SB_compa_all} 
\end{figure}

\subsubsection*{Comparison with Simulated bifurcation algorithm}
In order to perform comparative benchmarking of the BAWIM, we use the Simulated Bifurcation (SB) algorithm ~\cite{goto2019combinatorial, Ageron_Simulated_Bifurcation_SB_2023}, which has 
outperformed CIMs and simulated annealing in both solution quality and time-to-solution~\cite{goto2019combinatorial}. The SB algorithm has been further enhanced in subsequent works ~\cite{goto2021high, kanao2022simulated} by incorporating principles from classical mechanics and thermal heating. In this study, we use the Heated Ballistic Simulated Bifurcation (HbSB) variant of SB, as it has demonstrated improved performance among various versions of SB ~\cite{kanao2022simulated}.
Using HbSB, we conducted 1,000 simulations to determine the minimum Ising energy and the maximum MAX-CUT score, as summarized in Table \ref{tab:SBcompa}. Figure \ref{fig:SB_compa_all}(a) compares the time evolution of the best solution obtained by HbSB and BAWIM. Our machine required 341
ms to reach 90$\%$ of HbSB’s best Ising energy, whereas HbSB achieved the same energy in 102 
ms. The HbSB simulations were run on a modern system with an AMD Ryzen 7 9700X 8-core (5.30 GHz) CPU and 64 GB of RAM.
The performance of BAWIM can be significantly enhanced by utilizing a BAW delay line with a higher central frequency. In our current setup, we use a delay line with a 20.5 MHz central frequency, whereas BAW delay lines can operate at much higher frequencies up to 26 GHz. 
A catalog ~\cite{teledyne_BAW} from Teledyne lists the publicly available devices, among them, the MBJ-1018 device, operating at 16.455 GHz, can accommodate a similar number of spins to our present device. Operating the design at this higher frequency may require custom RF components as well as an FPGA capable of operating at higher frequencies.
Figure \ref{fig:SB_compa_all}(a) shows the projected runtime improvement using this 16.455 GHz delay line, where the target energy is estimated to be reached in just 0.462 ms, which is about 750 times faster than the current implementation.

Figure \ref{fig:SB_compa_all}(b) shows the success rate and time-to-target, relative to the best MAX-CUT score obtained by HbSB. Across all runs, BAWIM consistently achieves 99$\%$ of HbSB's optimal MAX-CUT score, with longer times required as the target increases. The BAWIM performance also varies with the problem size.

\subsubsection*{Dependence on the edge density}

\begin{figure}[!ht]
    \centering
    \includegraphics[width=0.99\linewidth]{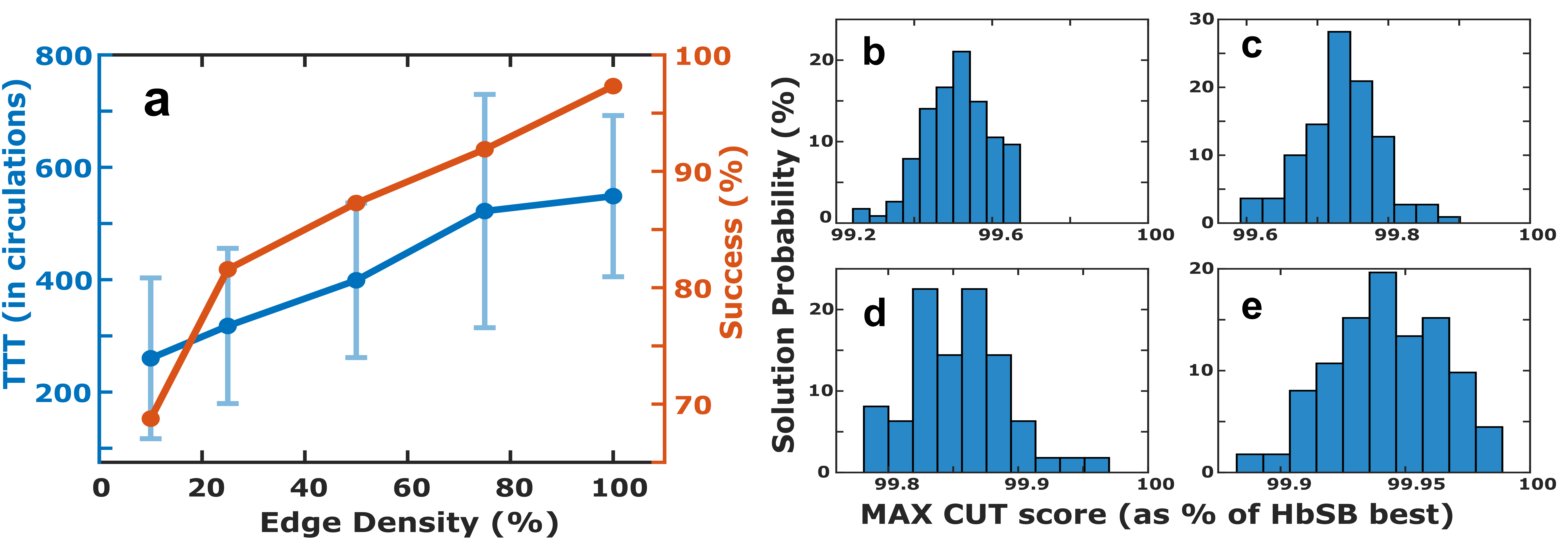}
\caption{
\textbf{(a) Time to solution and success rate for different edge density problems, with target Ising energy of $90\%$ of HbSB’s best result.}
The orange curve shows the success rate computed from n = 110 independent BAWIM runs. The blue curve shows the mean TTT computed over successful runs, and the error bars indicate $\pm1$ standard deviation.
\textbf{Solution probabilities for MAX-CUT problems with various edge densities}
\textbf{(b)} 25$\%$ edge density.
\textbf{(c)} 50$\%$ edge density.  
\textbf{(d)} 75$\%$ edge density.
\textbf{(e)} 100$\%$ edge density.
} 
\label{fig:EdgeDens_all} 
\end{figure}

To comprehensively benchmark our system, we evaluate its performance on problems with varying edge densities. Specifically, we test BAWIM using problems with edge densities of $25\%$, $50\%$, $75\%$, and $100\%$, generated from the BiqMac library. Figure \ref{fig:EdgeDens_all} presents the quantitative results based on over 100 solutions for each problem, showcasing the MAX-CUT scores in comparison to the HbSB's best result.
We also evaluated these problems using the HbSB algorithm. Table \ref{tab:SBcompa} reports the best observed Ising energy and MAX-CUT score from 1,000 simulations of the HbSB algorithm; it also includes the best performance achieved by our system among 100 runs. Here, we observe that BAWIM attains up to $95\%$ of HbSB’s best performance in terms of Ising energy, and up to $99.99\%$ in terms of the MAX-CUT score.

Figure \ref{fig:EdgeDens_all}(a) illustrates the time-to-target for problems with varying edge densities, using a target set at $90\%$ of HbSB’s best result. For the $10\%$ edge density problem, the time-to-target is approximately 341 ms and reaches the target $68\%$ of the time, whereas the fully connected ($100\%$ edge density) problem requires around 774 ms and achieves a success rate of $97\%$.

\begingroup
\setlength{\tabcolsep}{23pt} 
\begin{table}
    \centering
    \begin{tabular}{|c|c|c|}
    \hline
 \multirow{2}{6em}{Edge Density} &  \multicolumn{2}{c|}{Best Ising Energy}\\
         &  HbSB & BAWIM  \\
          \hline
        10$\%$ & -803,331,776  & $93.10 \%$  \\
         \hline
        25$\%$ & -1,201,604,480 & $93.59 \%$  \\
         \hline
         50$\%$& -1,491,890,816 & $95.52 \%$ \\
         \hline
        75$\%$ & -1,535,851,136 & $95.74 \%$\\
         \hline
         100$\%$ & -1,339,657,856 & $94.52 \%$\\
         \hline
    \end{tabular}
    \caption{Comparison of BAWIM and HbSB algorithms in terms of Ising energy across problems with different edge densities. The BAWIM Ising energies are expressed as percentages of the corresponding HbSB values.
    }
    \label{tab:SBcompa}
\end{table}
\endgroup

\subsection{Number Partitioning}
\begin{figure}[!ht]
    \centering
    \includegraphics[width=0.99\linewidth]{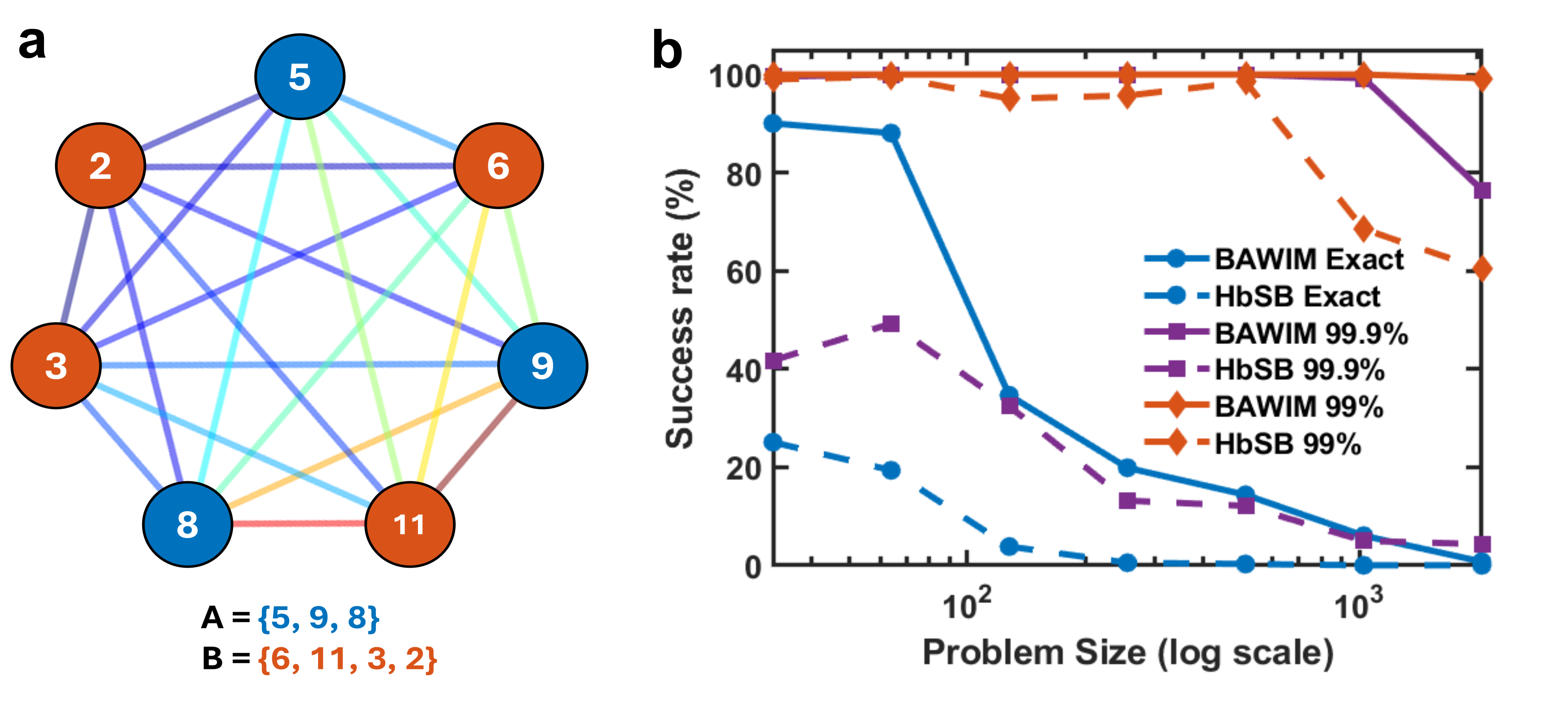} 
    \caption{ \textbf{Number partitioning problem}
    \textbf{(a)} Example of a number partitioning problem, represented as an all-to-all connected graph, where each spin denotes the assignment of a number to one of two subsets with equal sums.
    \textbf{(b)} The success rates of BAWIM and the HbSB algorithm are shown across problem sizes ranging from 32 to 2048, evaluated for different levels of approximate solution.} 
    \label{fig:Numpart_all}
\end{figure}

In this section, we focus on the number partitioning problem (NPP), which is one of Karp’s 21 NP-complete problems ~\cite{Karp1972}. It is defined as the task of dividing a given set of positive integers into two subsets such that the absolute difference between their sums is minimized, an example is shown in Fig. \ref{fig:Numpart_all}(a). 
The NPP has practical applications ranging from multiprocessor scheduling and minimizing the size and delay of the VLSI circuits ~\cite{coffman1991probabilistic, tsai1992asymptotic}, to public-key cryptography ~\cite{merkle1978hiding}, and in choosing sides in a ball game ~\cite{hayes2002computing}.
Furthermore, both the exact cover and the knapsack problems can be reduced linearly ~\cite{Karp1972, li2025efficient} to NPP without any constraints. This indicates that solving the NPP would also address these other NP-hard problems.

The NPP can be described as,
\begin{equation}
   E(\mathcal{A},\mathcal{B}) = \left| \sum_{i \in \mathcal{A}} a_i - \sum_{i \in \mathcal{B}} a_i \right|
\end{equation}
where $a_1, a_2,....a_N$ are the list of positive integers, and $\mathcal{A}$, $\mathcal{B}$ are the subsets of this list, and E represents the difference between the two groups, with E=0(1) representing the perfect partition when $\sum a_i$ is even(odd).

Let $s=+1$ represent subset $\mathcal{A}$ and $s=-1$ represent subset $\mathcal{B}$, then the above equation be written as
\begin{equation}
   E(s) = \left| \sum_{i=1}^{N} a_i s_i \right|
\end{equation}
\\
So, the coupling terms for the Ising model for the NPP can be written as $J_{ij}=a_i a_j$ with bias field $h=0$. This Hamiltonian for NPP is similar to the Mattis spin glass Hamiltonian~\cite{mattis1976solvable, grass2016quantum}.

To test the NPP, we generated seven distinct integer sets with sizes ranging from 32 to 2048, where each integer lies within the range of 1 to 178 and was generated randomly in Python. In BAWIM, the coupling elements $J_{ij}$ are limited to a 15-bit resolution, which limits the maximum integer values that can be used in the NPP.
These problem instances were evaluated using both the HbSB algorithm and the BAWIM. The results are summarized in Fig.~\ref{fig:Numpart_all}(b), here each set was tested for 10,000 runs with the HbSB algorithm and around 130 to 700 runs with the BAWIM. 
In BAWIM, smaller-node problems were run in parallel as replicas to get more quantitative results. In BAWIM, each solution was executed for 2 seconds, and the smallest difference obtained during each run was reported.
Here, the exact solution corresponds to a difference of 0, while the $99.9\%$ approximate solution indicates that the difference is less than $0.1\%$ of the maximum possible value (i.e., the total sum of all numbers). In all cases, BAWIM outperforms the HbSB algorithm, achieving a $100\%$ success rate in most instances when considering the $99.9\%$ approximate solution.

\subsection{Sudoku}
Sudoku is a famous puzzle game played on a $9 \times 9$ grid, where each cell can contain a number from 1 to 9. At the start, some cells are pre-filled with “clues,” while the rest remain blank. The goal is to complete the grid so that (i) every row, (ii) every column, and (iii) every 3 × 3 sub-grid (“block”) each contains all numbers from 1 to 9 exactly once.
The Sudoku puzzle originated in the 1970s as “Latin squares”, later adapted into its modern form with the name “Number Place”, and eventually rose to global popularity as “Sudoku” ~\cite{resnick2014sudoku, mucke2024simple}. Beyond attracting puzzle enthusiasts, it has also become a frequent subject of study for mathematicians and computer scientists ~\cite{resnick2014sudoku, pignari2025efficient, mucke2024simple, shukla2025non, ercsey2012chaos,lynce2006sudoku}, likely owing to its simplicity and accessibility to a non-technical audience.
For generalized grid sizes of $n^2 \times n^2$, the problem has been proven to be NP-complete~\cite{yato2003complexity}.

To adapt the Sudoku problem to the Ising model, we represent the digits 1 through 9 in each cell using a one-hot encoding scheme, since the Ising spins are inherently binary. This requires 9 spins per cell, leading to a total of $81 \times 9 = 729$ spins for the entire grid. This encoding adds a new rule that only one number should be present in each cell.
The coupling matrix is constructed by extending the one-hot encoding to incorporate the Sudoku rules, as outlined in ~\cite{mediumSUDOKUSolvingWorlds}. Clues are embedded in the bias field $h$, for a cell with a given clue, the bias of the corresponding spin is lowered while the biases of the other eight spins are raised, resulting in the clue being encouraged.

\begin{figure*}[!ht]
    \centering
    \includegraphics[width=0.95\linewidth]{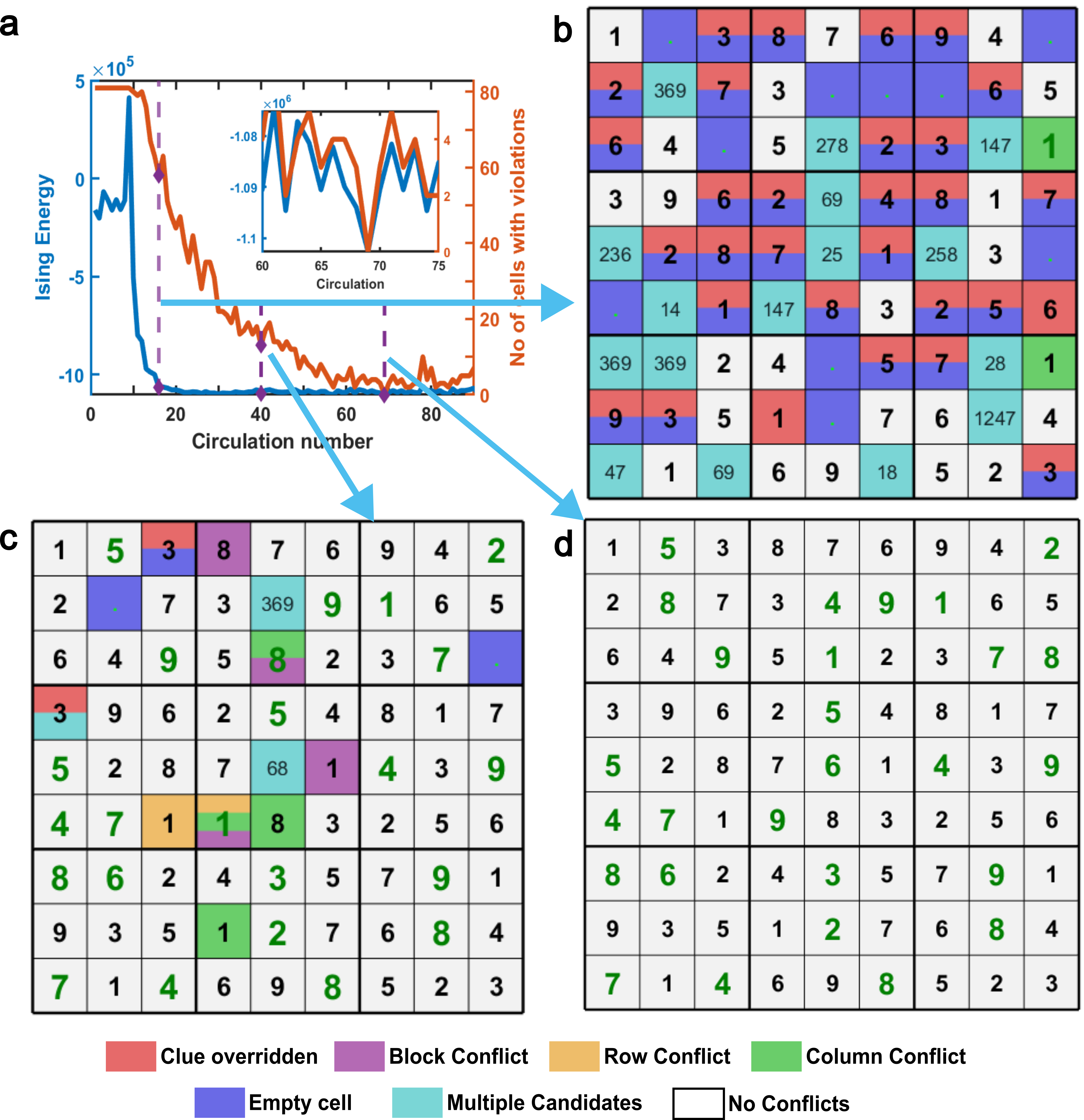}

\caption{\textbf{Sudoku puzzle}
\textbf{(a)} Evolution of the Ising energy and the number of cells with rule violations over time, the inset shows a zoomed-in view. The three vertical lines indicate the moments when the Ising energy reaches 96.71$\%$, 99.42$\%$, and 100$\%$ of the ground state. The corresponding Sudoku configurations are shown in
\textbf{(b)}, \textbf{(c)}, and \textbf{(d)}. The background colors highlight different types of issues associated with the cell values. The black numbers denote the original clues, multiple numbers within a single cell indicate multiple candidates violation, and the green numbers show the filled-in solutions.}
\label{fig:sudoku_ene_compa} 
\end{figure*}

Figure \ref{fig:sudoku_ene_compa}(a) illustrates the evolution of the Ising energy and the number of rule-violating cells during the Sudoku-solving process using BAWIM. Figures \ref{fig:sudoku_ene_compa}(b) and (c) display intermediate puzzle states corresponding to 96.71 $\%$ and 99.42 $\%$ of the ground-state energy, respectively. Notably, even when the Ising energy approaches the ground state, the Sudoku solution remains far from correct. The fully solved puzzle is shown in Fig. \ref{fig:sudoku_ene_compa}(d), where black numbers represent the given clues and green numbers indicate the completed solution. 
We also tested the Sudoku problems using the HbSB algorithm; however, it failed to produce correct solutions.
The Sudoku benchmark is included primarily to demonstrate the programmability of BAWIM for dense, non-native constraint-satisfaction problems. This highlights that, unlike other optimization problems where near-optimal states can still be useful, Sudoku requires exact constraint satisfaction, therefore, a near-ground-state Ising energy may still correspond to an invalid solution. 

\subsection{Estimation of Power consumption}
The total power consumption of BAWIM is approximately 9.57 W, primarily dominated by the FPGA and the five ZFL-500LN+ amplifiers. The AMD Zynq ZCU104 FPGA consumes about 5.2 W, with 2.40 W attributed to the programmable logic and 2.80 W to the processing system. The five ZFL-500LN+ amplifiers together draw 3.75 W, replacing them with specialized low-power amplifiers could reduce this to 50–100 mW. The remaining components, including the phase detector, RF switches, ADC, DAC, and multiplier, collectively contribute about 0.6 W.
By utilizing low-power amplifiers, the power consumption of BAWIM can be reduced to 5.9 W. 




\subsection{Comparison with CIMs}
The BAWIM builds on CIM foundations with major advances in functionality and architecture. Prior CIM implementations faced some limitations that our system avoids, for example, a 2000-spin CIM ~\cite{inagaki2016coherent} required extensive phase-locking and stabilization, restricting usable spins to 2048 out of 5082 DOPOs, with the rest devoted to auxiliary roles. In contrast, BAWIM uses 2048 of 2124 RF pulses directly as spins, with the remaining 76 left unused. Similarly, 100,000-spin CIM ~\cite{honjo2021100} also suffered from fiber-cavity phase instability, inconsistent solution times, and heavy post-selection, discarding up to half of the runs due to incomplete phase erasure. Our system operates stably without post-selection, improving both efficiency and throughput. Moreover, while CIM performance depended strongly on pump scheduling, BAWIM does not use such scheduling, yielding more predictable behavior.

Due to the disparity in carrier frequencies, BAWIM exhibits significantly higher efficiency in terms of carrier periods per spin and spin duty cycle. For comparison, the first CIM involved approximately $1\times 10^6$ carrier periods, while the more recent 100,000-spin CIM used around 6,000 
with a $15\%$ duty cycle. In contrast, BAWIM operates with just 6 carrier periods per spin and with a $50\%$ duty cycle.

In addition to the differences mentioned above, a significant challenge with CIMs is their sensitivity to temperature. The 100,000-spin CIM~\cite{honjo2021100} has a temperature coefficient of phase accumulation of $1.73 \times 10^7$ $deg/{^\circ} C$,  
meaning that a temperature change of just $1^\circ C$ can cause a shift in group delay of approximately 250 ps in a 5-km fiber delay line. To address this, CIMs must operate within a tightly controlled thermal environment with an accuracy of $\pm 0.05^\circ C$. The authors achieve this by placing the fiber inside a container with thick, hollow walls filled with water. Multiple thermistors and Peltier devices are used for active temperature regulation, and the setup is insulated with styrofoam and enclosed in an aluminum box.
In contrast, our BAWIM exhibits a temperature coefficient of phase accumulation of only 780 
${deg /^\circ}C$, which is $2.21 \times 10^4$ times lower than that of the CIM.
As a result, it requires no environmental or temperature control and can operate reliably at room temperature on a standard tabletop setup. 
Its ease of use, thermal stability, low power consumption, benchtop design, and the use of only off-the-shelf RF components make BAWIM a more advantageous and commercially viable option for combinatorial solvers.

\section{Conclusion}\label{conclu}
We presented the implementation of a time-multiplexed Ising machine based on bulk acoustic waves propagating through a quartz-based solid-state delay line. The BAWIM system features 2048 spins with 15-bit coupling resolution and with a circulation time of 1.41 ms.
We demonstrated its performance by solving the MAX-CUT problem on arbitrary graphs, achieving a solution time of 341 ms with the potential to reduce this to 0.46 ms. The BAWIM was benchmarked against graphs of varying densities from the BiqMac library.
Beyond MAX-CUT, we solve more complex tasks such as number partitioning and Sudoku, highlighting its practical utility for real-world problems.
Our design offers orders-of-magnitude higher thermal stability compared to state-of-the-art coherent Ising machines and surpasses the simulated bifurcation algorithm's performance in tackling complex problems.
Overall, the BAWIM provides a reliable, benchtop, thermally stable, and energy-efficient architecture with strong potential for further improvements in solution time and power consumption, with a promising prospect for commercialization.

\section{Methods}\label{methods}
\textbf{Sample: } The BAW delay medium is a fused-quartz disc shaped as a tetradecagon, measuring 9.5 cm across opposite sides. The microwave-to-acoustic wave transducers are made of Y-cut quartz crystal with silver electrodes.\\
\\
\textbf{Electrical measurements: } The frequency and group delay of the quartz delay line are derived from S-parameters measured using a vector network analyzer. Measurements were performed over a frequency range of 10–30 MHz with a resolution of 0.2 kHz.
Group delay is determined by taking the negative derivative of the phase shift with respect to frequency, and both the loss and time-delay curves are smoothed using a moving average filter with a window of 4000 points. 

The phase-sensitive amplifier was characterized by obtaining the envelope of an amplified signal with constant amplitude and a frequency detuned by 1 kHz from the reference signal. This detuning introduced a continuous phase drift with a 1 ms period, enabling the extraction of a smooth phase-sensitive amplification curve.


\section*{Acknowledgments}
This work was supported by a Knut and Alice Wallenberg Foundation WALP grant, KAW 253129326, a Horizon 2020 research and innovation program ERC advanced grant no. 835068 “TOPSPIN”, an ERC proof of concept grant no. 101069424 “SPINTOP”, and the Marie Skłodowska-Curie grant agreement no. 101111429 “SWIM”.

\section*{Author Contributions}
A.L. and J.\AA. conceived the concept. V.V. and A.L. designed the circuit. V.V performed the measurements and analyzed the data. V.V., R.O., V.G., and R.K. performed theoretical calculations. J.\AA. managed the project. All co-authors contributed to the manuscript, the discussion, and the analysis of the results.

\bibliography{References.bib}

@article{inagaki2016coherent,
  title={A coherent Ising machine for 2000-node optimization problems},
  author={Inagaki, Takahiro and Haribara, Yoshitaka and Igarashi, Koji and Sonobe, Tomohiro and Tamate, Shuhei and Honjo, Toshimori and Marandi, Alireza and McMahon, Peter L and Umeki, Takeshi and Enbutsu, Koji and others},
  journal={Science},
  volume={354},
  number={6312},
  pages={603--606},
  year={2016},
  publisher={American Association for the Advancement of Science}
}

@article{honjo2021100,
  title={100,000-spin coherent Ising machine},
  author={Honjo, Toshimori and Sonobe, Tomohiro and Inaba, Kensuke and Inagaki, Takahiro and Ikuta, Takuya and Yamada, Yasuhiro and Kazama, Takushi and Enbutsu, Koji and Umeki, Takeshi and Kasahara, Ryoichi and others},
  journal={Science advances},
  volume={7},
  number={40},
  pages={eabh0952},
  year={2021},
  publisher={American Association for the Advancement of Science}
}

@article{litvinenko2023spinwave,
  title={A spinwave Ising machine},
  author={Litvinenko, Artem and Khymyn, Roman and Gonz{\'a}lez, Victor H and Ovcharov, Roman and Awad, Ahmad A and Tyberkevych, Vasyl and Slavin, Andrei and {\AA}kerman, Johan},
  journal={Communications Physics},
  volume={6},
  number={1},
  pages={227},
  year={2023},
  publisher={Nature Publishing Group UK London}
}

@article{gonzalez2024global,
  title={Global biasing using a hardware-based artificial Zeeman term in spinwave Ising machines},
  author={Gonz{\'a}lez, Victor H and Litvinenko, Artem and Khymyn, Roman and {\AA}kerman, Johan},
  journal={Applied Physics Letters},
  volume={124},
  number={9},
  year={2024},
  publisher={AIP Publishing}
}

@article{gonzalez2024spintronic,
  title={Spintronic devices as next-generation computation accelerators},
  author={Gonz{\'a}lez, Victor H and Litvinenko, Artem and Kumar, Akash and Khymyn, Roman and {\AA}kerman, Johan},
  journal={Current Opinion in Solid State and Materials Science},
  volume={31},
  pages={101173},
  year={2024},
  publisher={Elsevier}
}

@article{litvinenko202550,
  title={A 50-spin surface acoustic wave Ising machine},
  author={Litvinenko, Artem and Khymyn, Roman and Ovcharov, Roman and {\AA}kerman, Johan},
  journal={Communications Physics},
  volume={8},
  number={1},
  pages={1--11},
  year={2025},
  publisher={Nature Publishing Group}
}

@article{takesue2025finding,
  title={Finding independent sets in large-scale graphs with a coherent Ising machine},
  author={Takesue, Hiroki and Inaba, Kensuke and Honjo, Toshimori and Yamada, Yasuhiro and Ikuta, Takuya and Yonezu, Yuya and Inagaki, Takahiro and Umeki, Takeshi and Kasahara, Ryoichi},
  journal={Science Advances},
  volume={11},
  number={7},
  pages={eads7223},
  year={2025},
  publisher={American Association for the Advancement of Science}
}

@article{barahona1982computational,
  title={On the computational complexity of Ising spin glass models},
  author={Barahona, Francisco},
  journal={Journal of Physics A: Mathematical and General},
  volume={15},
  number={10},
  pages={3241},
  year={1982},
  publisher={IOP Publishing}
}

@book{du1998handbook,
  title     = {Handbook of Combinatorial Optimization},
  editor    = {Du, Ding-Zhu and Pardalos, Panos M.},
  volume    = {4},
  year      = {1998},
  publisher = {Springer},
  address   = {New York, NY}
}

@article{ibarra1975fast,
  title={Fast approximation algorithms for the knapsack and sum of subset problems},
  author={Ibarra, Oscar H and Kim, Chul E},
  journal={Journal of the ACM (JACM)},
  volume={22},
  number={4},
  pages={463--468},
  year={1975},
  publisher={ACM New York, NY, USA}
}

@article{barahona1988application,
  title={An application of combinatorial optimization to statistical physics and circuit layout design},
  author={Barahona, Francisco and Gr{\"o}tschel, Martin and J{\"u}nger, Michael and Reinelt, Gerhard},
  journal={Operations Research},
  volume={36},
  number={3},
  pages={493--513},
  year={1988},
  publisher={INFORMS}
}

@article{earl2005parallel,
  title={Parallel tempering: Theory, applications, and new perspectives},
  author={Earl, David J and Deem, Michael W},
  journal={Physical Chemistry Chemical Physics},
  volume={7},
  number={23},
  pages={3910--3916},
  year={2005},
  publisher={Royal Society of Chemistry}
}

@article{vcerny1985thermodynamical,
  title={Thermodynamical approach to the traveling salesman problem: An efficient simulation algorithm},
  author={{\v{C}}ern{\`y}, Vladim{\'\i}r},
  journal={Journal of optimization theory and applications},
  volume={45},
  number={1},
  pages={41--51},
  year={1985},
  publisher={Springer}
}

@article{burke2004state,
  title={The state of the art of nurse rostering},
  author={Burke, Edmund K and De Causmaecker, Patrick and Berghe, Greet Vanden and Van Landeghem, Hendrik},
  journal={Journal of scheduling},
  volume={7},
  number={6},
  pages={441--499},
  year={2004},
  publisher={Springer}
}

@article{johnson2011quantum,
  title={Quantum annealing with manufactured spins},
  author={Johnson, Mark W and Amin, Mohammad HS and Gildert, Suzanne and Lanting, Trevor and Hamze, Firas and Dickson, Neil and Harris, Richard and Berkley, Andrew J and Johansson, Jan and Bunyk, Paul and others},
  journal={Nature},
  volume={473},
  number={7346},
  pages={194--198},
  year={2011},
  publisher={Nature Publishing Group UK London}
}

@inproceedings{nishiguchi2007single,
  title={Single-electron circuit for stochastic data processing using nano-MOSFETs},
  author={Nishiguchi, Katsuhiko and Fujiwara, Akira},
  booktitle={2007 IEEE International Electron Devices Meeting},
  pages={791--794},
  year={2007},
  organization={IEEE}
}

@article{mahboob2016electromechanical,
  title={An electromechanical ising hamiltonian},
  author={Mahboob, Imran and Okamoto, Hajime and Yamaguchi, Hiroshi},
  journal={Science advances},
  volume={2},
  number={6},
  pages={e1600236},
  year={2016},
  publisher={American Association for the Advancement of Science}
}

@article{sutton2017intrinsic,
  title={Intrinsic optimization using stochastic nanomagnets},
  author={Sutton, Brian and Camsari, Kerem Yunus and Behin-Aein, Behtash and Datta, Supriyo},
  journal={Scientific reports},
  volume={7},
  number={1},
  pages={44370},
  year={2017},
  publisher={Nature Publishing Group UK London}
}

@article{whitehead2023cmos,
  title={CMOS-compatible Ising and Potts annealing using single-photon avalanche diodes},
  author={Whitehead, William and Nelson, Zachary and Camsari, Kerem Y and Theogarajan, Luke},
  journal={Nature Electronics},
  volume={6},
  number={12},
  pages={1009--1019},
  year={2023},
  publisher={Nature Publishing Group UK London}
}

@article{si2024energy,
  title={Energy-efficient superparamagnetic Ising machine and its application to traveling salesman problems},
  author={Si, Jia and Yang, Shuhan and Cen, Yunuo and Chen, Jiaer and Huang, Yingna and Yao, Zhaoyang and Kim, Dong-Jun and Cai, Kaiming and Yoo, Jerald and Fong, Xuanyao and others},
  journal={Nature Communications},
  volume={15},
  number={1},
  pages={3457},
  year={2024},
  publisher={Nature Publishing Group UK London}
}

@article{yamamoto2017coherent,
  title={Coherent Ising machines—optical neural networks operating at the quantum limit},
  author={Yamamoto, Yoshihisa and Aihara, Kazuyuki and Leleu, Timothee and Kawarabayashi, Ken-ichi and Kako, Satoshi and Fejer, Martin and Inoue, Kyo and Takesue, Hiroki},
  journal={npj Quantum Information},
  volume={3},
  number={1},
  pages={49},
  year={2017},
  publisher={Nature Publishing Group UK London}
}

@article{goto2019combinatorial,
  title={Combinatorial optimization by simulating adiabatic bifurcations in nonlinear Hamiltonian systems},
  author={Goto, Hayato and Tatsumura, Kosuke and Dixon, Alexander R},
  journal={Science advances},
  volume={5},
  number={4},
  pages={eaav2372},
  year={2019},
  publisher={American Association for the Advancement of Science}
}

@misc{Ageron_Simulated_Bifurcation_SB_2023,
	author = {Ageron, Romain and Bouquet, Thomas and Pugliese, Lorenzo},
	title = {Simulated Bifurcation (SB) algorithm for Python},
howpublished = {https://github.com/bqth29/simulated-bifurcation-algorithm},
year = {2025}}

@article{goto2021high,
  title={High-performance combinatorial optimization based on classical mechanics},
  author={Goto, Hayato and Endo, Kotaro and Suzuki, Masaru and Sakai, Yoshisato and Kanao, Taro and Hamakawa, Yohei and Hidaka, Ryo and Yamasaki, Masaya and Tatsumura, Kosuke},
  journal={Science Advances},
  volume={7},
  number={6},
  pages={eabe7953},
  year={2021},
  publisher={American Association for the Advancement of Science}
}

@article{kanao2022simulated,
  title={Simulated bifurcation assisted by thermal fluctuation},
  author={Kanao, Taro and Goto, Hayato},
  journal={Communications Physics},
  volume={5},
  number={1},
  pages={153},
  year={2022},
  publisher={Nature Publishing Group UK London}
}

@Inbook{Karp1972,
author="Karp, Richard M.",
editor="Miller, Raymond E.
and Thatcher, James W.
and Bohlinger, Jean D.",
title="Reducibility among Combinatorial Problems",
bookTitle="Complexity of Computer Computations: Proceedings of a symposium on the Complexity of Computer Computations, held March 20--22, 1972, at the IBM Thomas J. Watson Research Center, Yorktown Heights, New York, and sponsored by the Office of Naval Research, Mathematics Program, IBM World Trade Corporation, and the IBM Research Mathematical Sciences Department",
year="1972",
publisher="Springer US",
address="Boston, MA",
pages="85--103",

isbn="978-1-4684-2001-2"}

@book{coffman1991probabilistic,
  title     = {Probabilistic Analysis of Packing and Partitioning Algorithms},
  author    = {Coffman, Edward Grady, Jr. and Lueker, George S.},
  year      = {1991},
  publisher = {Wiley\textendash Interscience},
  address   = {New York; Chichester}
}

@article{tsai1992asymptotic,
  title={Asymptotic analysis of an algorithm for balanced parallel processor scheduling},
  author={Tsai, Li-Hui},
  journal={SIAM Journal on Computing},
  volume={21},
  number={1},
  pages={59--64},
  year={1992},
  publisher={SIAM}
}

@article{merkle1978hiding,
  title={Hiding information and signatures in trapdoor knapsacks},
  author={Merkle, Ralph and Hellman, Martin},
  journal={IEEE transactions on Information Theory},
  volume={24},
  number={5},
  pages={525--530},
  year={1978},
  publisher={IEEE}
}

@article{hayes2002computing,
  title={Computing science: The easiest hard problem},
  author={Hayes, Brian},
  journal={American Scientist},
  volume={90},
  number={2},
  pages={113--117},
  year={2002},
  publisher={JSTOR}
}

@article{li2025efficient,
  title={Efficient solution of the number partitioning problem on a quantum annealer: a hybrid quantum-classical decomposition approach},
  author={Li, Zongji and Seidel, Tobias and Leib, Dominik and Bortz, Michael and Heese, Raoul},
  journal={Journal of Heuristics},
  volume={31},
  number={2},
  pages={21},
  year={2025},
  publisher={Springer}
}

@article{mattis1976solvable,
  title={Solvable spin systems with random interactions},
  author={Mattis, DC},
  journal={Physics Letters A},
  volume={56},
  number={5},
  pages={421--422},
  year={1976},
  publisher={Elsevier}
}

@article{grass2016quantum,
  title={Quantum annealing for the number-partitioning problem using a tunable spin glass of ions},
  author={Gra{\ss}, Tobias and Ravent{\'o}s, David and Juli{\'a}-D{\'\i}az, Bruno and Gogolin, Christian and Lewenstein, Maciej},
  journal={Nature communications},
  volume={7},
  number={1},
  pages={11524},
  year={2016},
  publisher={Nature Publishing Group UK London}
}

@techreport{resnick2014sudoku,
  title={Sudoku at the intersection of classical and quantum computing},
  author={Resnick, Timothy},
  year={2014},
  institution={Department of Computer Science, The University of Auckland, New Zealand}
}

@article{pignari2025efficient,
  title={Efficient solution validation of constraint satisfaction problems on neuromorphic hardware: the case of Sudoku puzzles},
  author={Pignari, Riccardo and Fra, Vittorio and Macii, Enrico and Urgese, Gianvito},
  journal={IEEE Transactions on Artificial Intelligence},
  year={2025},
  publisher={IEEE}
}

@inproceedings{mucke2024simple,
  title={A Simple QUBO Formulation of Sudoku},
  author={M{\"u}cke, Sascha},
  booktitle={Proceedings of the Genetic and Evolutionary Computation Conference Companion},
  pages={1958--1962},
  year={2024}
}

@article{shukla2025non,
  title={Non-binary dynamical Ising machines for combinatorial optimization},
  author={Shukla, Aditya and Erementchouk, Mikhail and Mazumder, Pinaki},
  journal={Physica D: Nonlinear Phenomena},
  pages={134809},
  year={2025},
  publisher={Elsevier}
}

@article{ercsey2012chaos,
  title={The chaos within Sudoku},
  author={Ercsey-Ravasz, M{\'a}ria and Toroczkai, Zolt{\'a}n},
  journal={Scientific reports},
  volume={2},
  number={1},
  pages={1--8},
  year={2012},
  publisher={Nature Publishing Group}
}

@inproceedings{lynce2006sudoku,
  title={Sudoku as a SAT Problem.},
  author={Lynce, In{\^e}s and Ouaknine, Jo{\"e}l},
  booktitle={AI\&M},
  year={2006}
}

@article{yato2003complexity,
  title={Complexity and completeness of finding another solution and its application to puzzles},
  author={Yato, Takayuki and Seta, Takahiro},
  journal={IEICE transactions on fundamentals of electronics, communications and computer sciences},
  volume={86},
  number={5},
  pages={1052--1060},
  year={2003},
  publisher={The Institute of Electronics, Information and Communication Engineers}
}

@misc{mediumSUDOKUSolvingWorlds,
	author = {Yuichiro Minato},
	title = {{S}olving the world’s most difficult {S}{U}{D}{O}{K}{U} problem using ising model on javascript --- minatoyuichiro.medium.com},
	howpublished = {https://minatoyuichiro.medium.com/sudoku-solving-the-worlds-most-difficult-sudoku-problem-using-ising-model-on-javascript-c9b5add0e5c0},
	year = {}}

@article{helmberg2000spectral,
  title={A spectral bundle method for semidefinite programming},
  author={Helmberg, Christoph and Rendl, Franz},
  journal={SIAM Journal on Optimization},
  volume={10},
  number={3},
  pages={673--696},
  year={2000},
  publisher={SIAM}
}

@misc{rudy_gen,
	author = {Giovanni Rinaldi},
	title = {rudy graph generator},
	howpublished = {www-user.tu-chemnitz.de/˜helmberg/rudy.tar.gz},
	year = {1996}}

@misc{teledyne_BAW,
	author = {Teledyne},
	title = {BAW Devices Product Selection Guide},
	howpublished = {https://www.teledynedefenseelectronics.com/wireless/Wireless Brochures/BAW Devices Product Selection Guide.pdf},
    year = {}}

@article{ovcharov2024numerical,
  title={A numerical model for time-multiplexed Ising machines based on delay-line oscillators},
  author={Ovcharov, Roman V and Gonz{\'a}lez, Victor H and Litvinenko, Artem and {\AA}kerman, Johan and Khymyn, Roman S},
  journal={arXiv preprint arXiv:2406.07197},
  year={2024}
}

@book{hashimoto2009rf,
  title     = {{RF} Bulk Acoustic Wave Filters for Communications},
  editor    = {Hashimoto, Ken-ya},
  year      = {2009},
  publisher = {Artech House},
  address   = {Norwood, MA}
}

@article{liu2020materials,
  title={Materials, design, and characteristics of bulk acoustic wave resonator: A review},
  author={Liu, Yan and Cai, Yao and Zhang, Yi and Tovstopyat, Alexander and Liu, Sheng and Sun, Chengliang},
  journal={Micromachines},
  volume={11},
  number={7},
  pages={630},
  year={2020},
  publisher={MDPI}
}

@article{tang2024review,
  title={A review of surface acoustic wave sensors: Mechanisms, stability and future prospects},
  author={Tang, Zhaozhao and Wu, Wenyan and Yang, Po and Luo, Jingting and Fu, Chen and Han, Jing-Cheng and Zhou, Yang and Wang, Linlin and Wu, Yingju and Huang, Yuefei},
  journal={Sensor Review},
  volume={44},
  number={3},
  pages={249--266},
  year={2024},
  publisher={Emerald Publishing Limited}
}

@article{mandal2022surface,
  title={Surface acoustic wave (SAW) sensors: Physics, materials, and applications},
  author={Mandal, Debdyuti and Banerjee, Sourav},
  journal={Sensors},
  volume={22},
  number={3},
  pages={820},
  year={2022},
  publisher={MDPI}
}

@inproceedings{aigner2018baw,
  title={BAW filters for 5G bands},
  author={Aigner, R and Fattinger, G and Schaefer, M and Karnati, K and Rothemund, R and Dumont, F},
  booktitle={2018 IEEE International Electron Devices Meeting (IEDM)},
  pages={14--5},
  year={2018},
  organization={IEEE}
}

@article{terzieva2016overview,
  title={Overview of bulk acoustic wave technology and its applications.},
  author={Terzieva, Milena D},
  journal={Electrotechnica \& Electronica (E+ E)},
  volume={51},
  year={2016}
}
\end{document}